# Transforming how water is managed in the West

Patrick Atwater
Project Manager
California Data Collaborative
patrick@argolabs.org

Christopher Tull
Civic Data Scientist
California Data Collaborative
Ctull17@gmail.com

Eric Schmitt
Advising Statistician
California Data Collaborative
Eric.Schmitt@protix.eu

Joone Lopez
Lead Administrator
California Data Collaborative
JLopez@mnwd.com

Drew Atwater
Operational Administrator
California Data Collaborative
DAtwater@mnwd.com

Varun Adibhatla
Head of Rapid Prototyping
California Data Collaborative
varun@argolabs.org

## ABSTRACT

California is challenged by its worst drought in 600 years and faces future water uncertainty. Pioneering new data infrastructure to integrate water use data across California's more than a thousand water providers will support water managers in ensuring water reliability. The California Data Collaborative is a coalition of municipal water utilities serving ten percent of California's population who are delivering on that promise by centralizing customer water use data in a recently completed pilot project. This project overview describes tools that have shown promising early results in improving water efficiency programs and optimizing system operations. Longer term, these tools will help navigate future uncertainty and support water managers in ensuring water reliability no matter what the future holds. The uniquely publicly-owned data infrastructure deployed in this project is envisioned to enable the world's first "marketplace of civic analytics" to power the volume of water efficiency measurements water managers require at a radically more cost effective price. More broadly, this data-utility approach is adaptable to domains other than water and shows specific potential for the broader universe of natural resources.

## 1.INTRODUCTION

Customer meter level water use data may seem arcane and abstract, yet it is telling that the White House, a lead March 17, 2016 New York Times Op-Ed, the Western Governors Association, the California Council on Science and Technology and others have called for precisely the water data integration this project is delivering. Since launching in January 2016, the California Data Collaborative has already been featured as part of the White House Water Summit on March 22, 2016 and pioneered a new path for smart statewide efficiency.

The California Data Collaborative coalition of water utilities serves 3.7 million Californians directly through retail water service and serves over half of the state on a wholesale level. This group of water managers came together out of a shared frustration with suboptimal water us e data collected by the state that compares local utilities using imperfect, one-size-fits-all averages. Not content to simply accept the status quo, the California Data Collaborative has piloted new data infrastructure and centralized actual metered water use to power a quantum leap in smart water efficiency.

The data infrastructure developed by project staff aligns disparate utility-specific customer codes with standard statewide customer classifications along with key contextual information like evapotranspiration, irrigable area, and demographics. This standardized data can benchmark water efficiency through meaningful comparisons using California's existing Model Water Efficient Landscape Ordinance (MWELO) and support water managers in actually achieving that efficiency by providing tools to measure past performance and target future water efficiency goals.

The California Data Collaborative has piloted new data infrastructure to centralize that important customer water use data, leveraging open source modern data ingestion and cloud data warehouse tools. This platform is called the Strategic California Urban Water Use data infrastructure, or "SCUBA". Critically, this data is standardized into general customer and water use categories across utilities. SCUBA:

- Aligns with existing California state Department of Water Resources classifications (Single Family Residential, Commercial, etc.)

- Distinguishes water supply type ("Potable," "Recycled," and "Raw" Water)

- Provides single and master metered flags to distinguish apartments

- Integrates contextual demographic information through county assessor and census identifiers

- Links to conservation actions undertaken by utilities

demand forecasting. The following sections detail each of those aspects in turn.

## 2.DEVELOPING AN INTEGRATED SUITE OF WATER EFFICIENCY ANALYTICS

*2.1. Statistical estimates of the water savings associated with turf removal*

The Data Collaborative has applied statistical techniques to estimate the actual water savings achieved through turf removal rebates. When this is combined with automated monthly data ingestion, it provides the ability to generate ongoing estimates of program savings to adaptively manage the historic investment in turf rebates. This sort of approach also opens the door to more advanced analyses such as outlier detection of accounts that do not achieve the desired savings level.

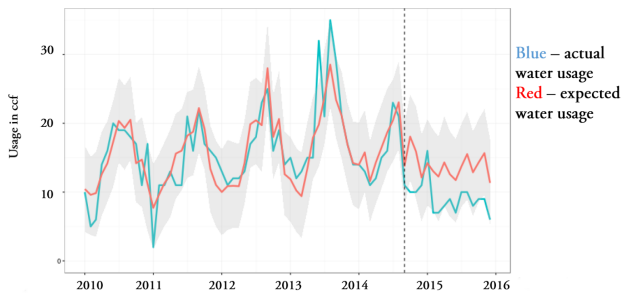

*Figure 1. The blue line shows the observed usage for one household. The red line shows the expected usage based on the consumption of households with historically similar usage patterns. The vertical dotted line represents the time the turf removal was performed. A 29 percent reduction from expected water use is visible after the removal.*

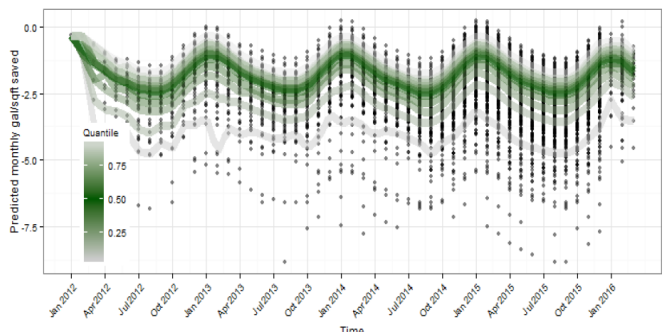

*Figure 2. Predicted monthly savings for each household in the data set. The dark green line corresponds to median savings. Seasonal variation leads to swings in average savings from -1.5 to -2.7 gallons per square foot.*

Mean predicted savings for single-family residential accounts are estimated at 24.6 gallons per square foot per year for the households used in this study (Tull, Schmitt, Atwater 2016). This translates into a present value cost of $320.02 per acre foot of water saved, assuming a ten-year lifespan and a hyperbolic discount rate. That cost decreases to $142.21 assuming a thirty-year generational lifespan.

The study utilizes a data set of 545 unique single-family residential turf rebates across 3 California water utilities, totaling 635,713 square feet of converted turf grass to estimate the water savings from turf removal. More broadly, the approach can be applied to estimate the causal impact of any targeted conservation program by comparing the usage patterns of affected and unaffected customers and how these patterns change after a program takes effect.

STATUS: This method has been presented to our technical working group on May 17, 2016 and is currently under peer review through our internal Data Collaborative processes and externally through acceptance into the KDD Workshop on Data Science for Food, Energy and Water 2016.

*2.2. Assessing landscape area changes using publicly available aerial imagery*

The preceding turf rebate participation dynamics and water savings analyses omit a key data point: the adoption rate of California-friendly landscaping of households that did not take a rebate. Tracking changes in land cover over time can yield insights into the effectiveness of both turf removal rebates and general trends towards the transformation of outdoor landscapes. The Data Collaborative is working with Professor Andrew Marx of the Claremont Graduate University to generate accurate, automated measurements of land cover classifications using free imagery available through the National Agriculture Imagery Program (NAIP). NAIP imagery is updated every two years at a resolution of 1 meter for the state of California. When combined with algorithms to automatically classify ground cover, this will enable two key measurements:

1. Changes in land cover over time to determine peer effects of turf rebates and progress towards landscape transformation
2. Aggregate utility level irrigable area land cover analysis to support meaningful benchmarking of efficient water use

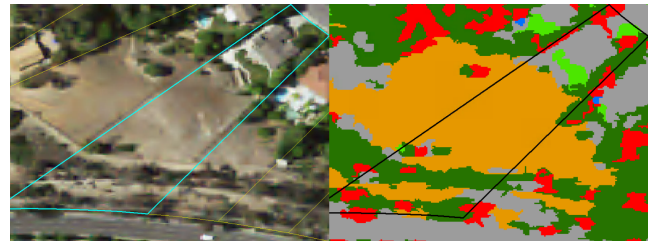

*Figure 3. Classification of ground cover types from free NAIP imagery. Left shows an unaltered NAIP image with parcel polygon. Right shows the classifier output with colors representing different ground covers. Turfgrass is visible as bright green.*

STATUS: Andrew Marx has piloted this approach pro-bono in Mission Viejo and has been contracted by the Data Collaborative to deploy this methodology across Metropolitan Water District's service area. Data Collaborative staff have experimented with the Google Earth Engine to scale this methodology statewide. In addition, staff has been developing an open assessment of vendor accuracy to provide a "consumer" reports function and objectively evaluate various landscape area classification providers.

*2.3. Deploying econometric models to measure the impact of price shifts on water sales*

Data Collaborative staff have been working closely with several members of the technical working group to scope a new rate modeling tool that can be deployed in approximately six months. This tool will help water managers analyze how many customers would fall into each proposed usage tier according to historical data and how much revenue that model would generate.

The next step is to model how a price shift impacts water sales. To this end, the Collaborative has partnered with Dr. Kenneth Baerenklau and the Water Science and Policy Center at UC Riverside to routinize best-in-class econometric analyses to estimate the impact of rate shifts on water demand. Dr. Baerenklau has previously demonstrated the ability of block rate water budgets to decrease water demand (Baerenklau, Schwabe, Dinar 2014), and is providing assistance in adjusting his models for application within the Collaborative. Streamlining econometric estimation of the price elasticity of water demand will help contextualize how different communities with different demographics and set of other water conservation actions respond to rate shifts. Ultimately, routinizing this form of econometric study will help predict the effect of rate changes on water sales for any utility that joins the Collaborative and has their data integrated into the SCUBA data warehouse.

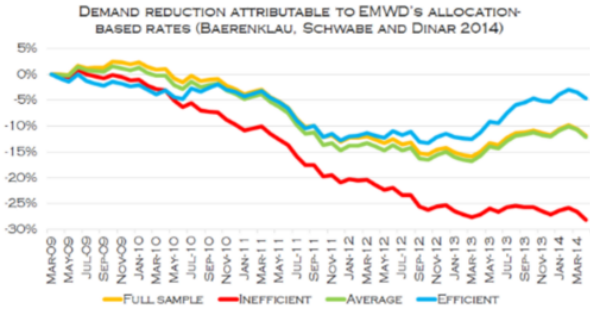

*Figure 4. UC Riverside econometric evaluation of the impact of Eastern Municipal Water District rate shift on water demand*

STATUS: Data Collaborative staff has been conducting an extensive literature review and working with rate experts on a roadmap discussed at the June 8th Technical Working Group meeting at IRWD. In addition, the UCR economists have undertaken an ongoing study for MNWD district which notably includes a census of single family customer attitudes.

*2.4. Leveraging randomized control trials to measure marketing effectiveness*
The Collaborative is working with Stanford Professor Wes Hartman on two approaches to measure the effectiveness of conservation marketing over social media. The first tests the efficacy of educational and action-oriented messaging by randomly assigning those treatments to households. Another program evaluates the effects of smart water controllers in a randomized trial whereby customers are given access to a vendor's micro-site where they get prices that are highly discounted based on preferred pricing and rebates the utility may offer.

STATUS: Data Collaborative staff has helped facilitate additional utilities participating in Wes Hartman's ongoing experiments. The first round of participation occurred in May 2016 and another round is anticipated in the Fall. In addition, the Data Collaborative has worked with members of the technical working group to develop standardized mechanisms to track marketing and outreach across agencies in order to routinize this sort of evaluation of conservation marketing effectiveness.

*2.5. Water demand forecasting and improving system operations*
A more robust understanding of water demand can also inform improved water system operations. This forms part of an integrated vision that we articulated in the feature JAWWA article in June 2015 on how the sort of data tools articulated in this prospectus can improve water management in the twenty first century.

"Extreme weather and uncertain future water supplies challenge utilities to do more to integrate resources and adapt to a rapidly changing world. New digital data management and analysis tools allow utilities to better share information as well as regularly model and update future water resource projections of supplies and demand. This iterative approach allows utilities to proactively navigate an uncertain future."

"How Water Utilities Can Plan for Uncertainty,"
Richard Atwater, Patrick Atwater, Drew Atwater, Johnathan Cruz

The California Data Collaborative is currently working with a team of DataKind volunteers to test and deploy short term demand forecasting tools to improve recycled water operations. With accurate 1-4 week short term demand forecasting, California's water managers might better adjust the timing and routing of water through their network to plan for peak needs, schedule the movement of water around peak electric prices and minimize water lost to pressure and evaporation.

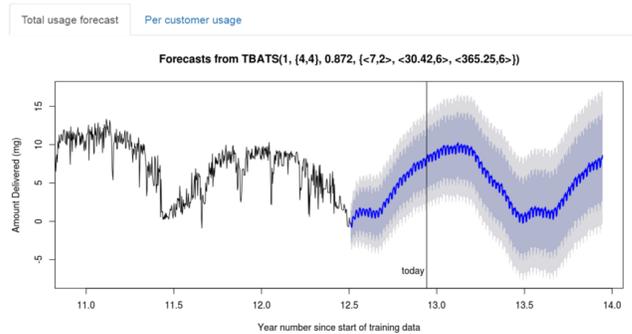

*Figure 5. This interactive time series tool shows preliminary demand forecasting results from a collaboration with DataKind, a nonprofit organization back by the Clinton Global Initiative that brings talented volunteer data scientists to bear on important public problems.*

STATUS: The Data Collaborative Project Manager Patrick Atwater and MNWD day to day lead administrator Drew Atwater have been meeting weekly with a team of pro-bono data scientists led by a senior Netflix data scientist.

*2.6. Optimizing system operations and water / energy efficiency*
This improved understanding of demand can also connect to improved operations through the water / energy nexus. In California, the state collects over one billion dollars annually as part of a public goods surcharge on energy. Much of this money is invested into energy efficiency programs. Recently, there has been interest in the state to invest some of these funds into water use efficiency and capital improvements to water infrastructure. However, before such investments can be made, energy utilities need to understand how much energy is saved through improvements to water systems. Energy is used to source, convey, treat, use, and dispose of or recycle water (Escriva-Bou 2014). Data collected across the entire lifecycle of water are fragmented across multiple databases, with varying levels of concern over security and privacy. Seamless integration of these data could enable energy utilities to invest in the water sector. Additionally, with an understanding of the energy savings associated with water use efficiency programs and capital improvements to water infrastructure, water and energy utilities may be eligible for additional funds through the California Carbon Cap-and-Trade market.

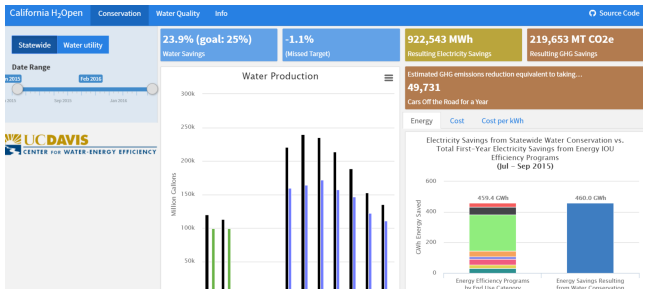

*Figure 6. This application uses the reported data to visualize how different water utilities have responded to this mandate. In addition to displaying a summary of water use relative to the conservation target for each district, we also calculate the electricity savings associated with the reduced demand on water infrastructure services using estimates of average energy intensity per hydrologic region. We then convert the electricity savings into avoided greenhouse gas (GHG) emissions based on the emissions factor specific to the water utility's regional electricity provider. For more detail please check out:*
*https://cwee.shinyapps.io/greengov/*

STATUS: UC Davis CWEE has developed a dashboard quantifying the embedded energy and greenhouse gas emissions avoided through water conservation. This dashboard was presented to the Water Energy Team of the Climate Action Team ("WET CAT"). Plans to improve those utility level dashboards with granular customer data like that managed by the California Data Collaborative are currently under development.

## 3. PIONEERING THE WORLD'S FIRST "MARKETPLACE OF CIVIC ANALYTICS"

California's drought has attracted worldwide attention, and the SCUBA data infrastructure provides a path to bring the best and brightest analytical minds to bear on achieving the water efficiency our state needs. Our vision is to develop a "marketplace of analytics" whereby any qualified researcher or analyst from anywhere on the globe can offer to provide the water efficiency analytics California water managers need.

The current ad hoc water efficiency analytical environment involves unnecessarily high transaction costs through 1) bespoke legal agreements 2) differing data standards and formats 3) lack of a centralized clearinghouse to vet prospective analysts. These high transaction costs largely prevent all but the most sophisticated utilities from participating in the current water efficiency research marketplace.

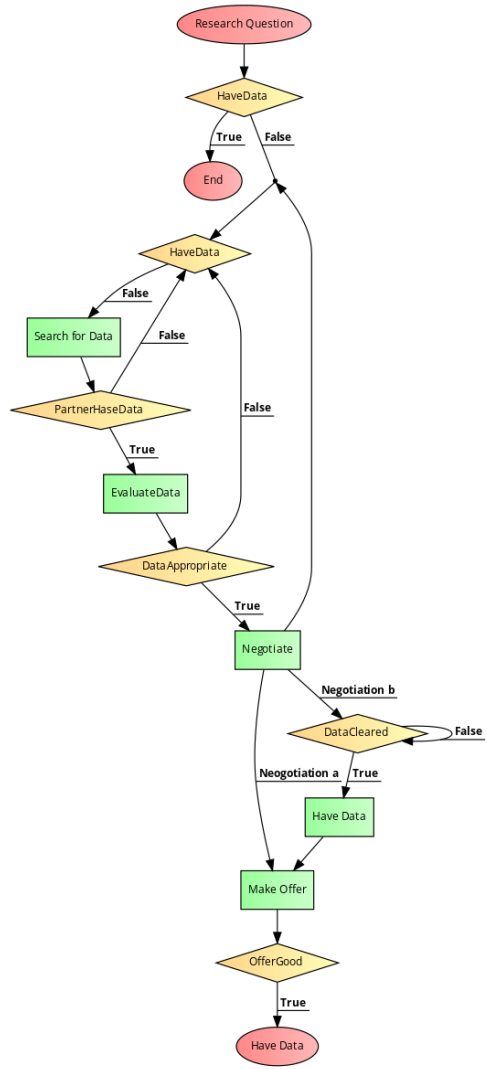

*Figure 7. An information process diagram for the current steps involved in searching for water data for a research project. This process involves unnecessarily high transaction costs that precludes many researchers and water agencies from participating in the "marketplace" of water research.*

Centralizing high value water data like meter level customer use into standardized statewide customer classifications provides radically more streamlined technical logistics for researchers and analysts to develop analytics. Coupled with a robust legal framework to streamline how utilities greenlight resharing data, this approach will power the water efficiency analytics measuring the impact of rebates, rates and other actions detailed in the above sections that water managers need to navigate the new normal of increased water scarcity.

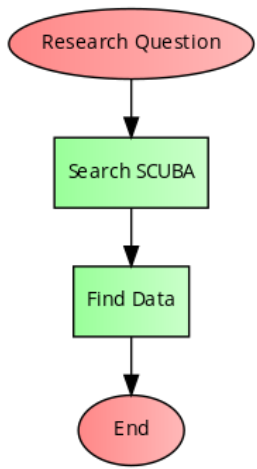

*Figure 8. An information process diagram illustrating how the SCUBA data infrastructure streamlines the search process for California urban customer use water data.*

In addition, this approach has significant research benefits and can power categorically more robust analyses. The economic blog Interfluidity outlines how this new, digitally native approach to social science would enable researchers and analysts to:

*"...build new analyses from any stage of old work, by recruiting raw data into new projects, by running alternative models on already cleaned-up or normalized data tables, by using an old model's estimates to generate inputs to simulations or new analyses." (Interfluidity 2014)*

That marketplace of ideas approach would integrate the best of the software analytics' scale, academic researchers' depth and professional services' ability to customize services to utilities' needs. The California Data Collaborative has partnered with a wide spectrum of academic researchers and multidisciplinary partnerships to develop and nurture this nascent ecosystem of water efficiency analytics.

## 4. PROJECT ADMINISTRATION

This uniquely water manager led "Big Data" project brings together utilities across California to collaborate by sharing data and technical expertise. As of June 2016, the following utilities are project partners:

- ● *Eastern Municipal Water District* provides retail water service to approximately 142,000 connections and imports water for 785,000 people in Riverside County.
- ● *East Bay Municipal Services District* provides water and wastewater services for 1.4 million customers in Alameda and Contra Costa counties.
- ● *Las Virgenes Municipal Water District* provides potable water, wastewater treatment, recycled water and biosolids composting to more than 65,000 residents in Western Los Angeles County.
- ● *Inland Empire Utilities Agency* provides wastewater and wholesale water services to 830,000 people in Western San Bernardino County.
- ● *Irvine Ranch Water District* provides water, wastewater collection and treatment, recycled water programs, and urban runoff treatment to more than 380,000 residents in Orange County.
- ● *The Metropolitan Water District of Southern California* is a regional wholesaler that delivers water to 26 member public agencies – 14 cities, 11 municipal water districts, one county water authority – which in turn provides water to more than 19 million people in Los Angeles, Orange, Riverside, San Bernardino, San Diego and Ventura counties.
- ● *Monte Vista Water District* provides retail and wholesale water service to over 130,000 people in Montclair, Chino Hills, portions of Chino and the unincorporated area lying between the cities of Pomona, Chino Hills, Chino and Ontario.
- ● *Moulton Niguel Water District* provides water, wastewater, and recycled water services to 170,000 people in South Orange County.
- ● *Santa Margarita Water District* provides water and wastewater services to over 155,000 customers in Mission Viejo, Rancho Santa Margarita and the unincorporated areas of Coto de Caza, Las Flores, Ladera Ranch and Talega.

This group of utilities serves approximately 3.7 million Californians through retail water service and approximately 21 million people through wholesale water service.

Sharing data across these utilities is made possible by utility staff participating in the technical working group and following project staff members of Team ARGO, interns and research fellows.

**Project Manager**
Patrick Atwater has over five years of experience in data intensive roles in the water and information technology industries. He ran the numbers for the State Water Contractors on Governor Brown's $13 billion Bay Delta fix and co-authored the feature June 2015 American Water Works Association article on how data science can help water utilities adapt to climate change. Projects include: developing the financial model for the proposed Claremont Colleges 300,000 gallons per day MBR recycled water plant; generating financial analysis at 14 local government agencies with combined budget of over $5 billion; CFO level analysis for a first-of-its-kind securitization structure for a $700 million groundwater treatment facility; and presenting California's first 2010 redistricting GIS analysis to an audience of over 100 municipal managers.

**Civic Data Scientist**
Christopher Tull recently graduated with a Masters from the Center for Urban Science and Progress, the nation's leading civic data science program founded by ex-Cal Tech provost Steven Koonin and launched as part of the Bloomberg applied sciences initiative. He has worked as a research assistant at the Max Planck Institute for Biological Cybernetics as well as NYU CUSP, where his work on energy-water usage intensity in NY won best paper at the Bloomberg Data for Good Exchange. He is highly proficient in the following programming languages: Python, R, Java, Javascript, C++, C, SQL. In addition, Chris has experience working with the following technologies: ArcGIS, CartoDB, Git, Eclipse, RStudio, IPython Notebook, Excel, Mysql, Postgresql/PostGIS.

**Consulting Statistician**
Eric Schmitt co-authored the MNWD turf rebate study published by Bloomberg Data for Good in September 2015. Previously, Eric has worked as an economic consultant at NERA Economic Consulting for two years performing econometric analyses on topics ranging from anti-trust to consumer behavior in the tobacco industry. He has completed a PhD in statistics from KU Leuven University and currently works as a statistician at Protix Biosystems, a pioneering technology company creating efficiencies in the water-food nexus through high volume insect products. He has provided statistical consulting services for and spoken before leading companies, such as Mars, Inc. and Tableau Technologies, and is the author of peer-reviewed articles in theoretical and industrial statistics, education, and medicine.

**Head of Rapid Prototyping**
Varun Adibhatla has spent the past 10 years studying and implementing technology in domains ranging from decision support systems for crisis response to supporting high frequency & algorithmic trading at large banks. Varun is a founding member of ARGO Labs, which advocates civic data science, and developed the Street Quality Identification Device, a recipient of the Knight Foundation's Prototype grant to enable low-cost, citywide digital surveys of street quality. He has a Masters in Human Computer Interaction as well as Urban Science and Informatics.

**Data Systems Engineer**
Graham Henke is a computer science graduate from Purdue University, Graham worked at Apple Inc for 5 years before setting his sights on NYC. As a recent graduate from NYU CUSP, Graham is now focused on how technology can be used to address challenges faced by cities. He has full stack engineering skills and has experience managing back end data infrastructure for ARGO's street quality identification device project.

**Front End Data Scientist**
David Marulli graduated with a Masters in Urban Science and Informatics from NYU CUSP. Post CUSP he worked for the Rudin Center developing transportation focused data visualizations. He is skilled in quantitative methods, modern programming tools and working with municipal government administrative data.

**System Architecture Research Fellow**
Tony Castalletto is a versatile technology expert with over a decade of experience leading and implementing complex projects for leading educational and research institutions. Innovative problem solver with a long track record developing robust solutions to technology problems.

**Urban Water Efficiency Research Fellow**
Brianna Pagan studies climate change impacts on the hydrological cycle in the Western United States and works to provide practical solutions to potential adverse effects on water supply. Uniting her background in water policy and environmental science, Brianna's goals are to further understand the potential impacts of climate change and effectively integrate her research into state and local water agencies' long term planning strategies.

**Public Affairs Intern**
Wendy Greene obtained her Masters in Science degree from UCLA in Environmental Health Sciences and has been involved in environmental leadership, programming, and policy throughout her academic career. She has experience in leadership roles in a half dozen UCLA natural resource and sustainability student organizations.

ARGO's work to power the California Data Collaborative has also included a novel partnership with Enigma Technologies, a leading civic data startup that powers the world's largest repository of public data. The tools enabling the integration of that data have been provided pro-bono by Engima to scale the California Data Collaborative's early success to realize its vision of integrating the entire lifecycle of water data in California and beyond. California Governor Brown has called for bringing this sort of "pioneering spirit" into California government since the start of his second stint as governor, a vision this project aims to deliver.

*"The world has changed far faster than government's ability to keep pace, creating a huge space for good government reforms to better society. In William Mulholland's era, Los Angeles could get its water through the work of a single agency acting essentially in isolation. Today, however, not only do you need coordination between multiple agencies at multiple levels of government that simply deal with water, but our world is fundamentally more connected, with profound institutional consequences.*

*Operating that water infrastructure is predicated on a vast array of telecommunications and electrical systems, involving several more sets of public and private actors. Even NASA and the military are involved. Refurbished predator drones are flown over the Bay Delta to gather environmental quality data. Today a dazzling array of interlocking parts work together to ensure Californians have a clean, secure, and sustainable supply of water.*

*The fundamental challenge California faces – getting water from where it falls to where it's needed – hasn't changed. But rather than having a set of institutions designed to solve that problem, we've settled for a byzantine structure that only exists because that's the way things have always been. So why not unleash the famed creativity of the California people to systematically rethink how government can address the fundamental challenges – schools, prisons, water, public safety, etc. – we face as a people."*

– "A New California Dream" by Patrick Atwater

## 5. ACKNOWLEDGEMENTS

The authors would like to thank our utility partners including Moulton Niguel, Irvine Ranch, Eastern Municipal, Las Virgenes Municipal, Santa Margarita, and Monte Vista water districts, along with the Inland Empire Utilities Agency, the East Bay Municipal Utility District, and the Metropolitan Water District of Southern California, Enigma Technologies, Facebook, OmniEarth, UC Davis Center for Water and Energy Efficiency, Seed Consulting Group, Microsoft, DataKind, Stanford, University of California Riverside, Los Angeles, Netflix, Claremont Graduate University, the California Urban Water Conservation Council, the Association of California Water Agencies, the Southern California Water Committee and Cal Tech for all of their support for and participation in the California Data Collaborative.